%% file: provenance.tex
\documentclass[runningheads]{llncs}

\usepackage{xcolor}
\usepackage{hyperref}
\usepackage{graphicx}
\usepackage{balance}  % for  \balance command ON LAST PAGE  (only there!)
\usepackage{longtable}
\usepackage{tabu}

\usepackage{amsmath}
\usepackage{bm}
\usepackage{amssymb}
\usepackage{algorithm}
\usepackage[noend]{algpseudocode}
\usepackage{verbatim}
\usepackage{array}
\usepackage{multicol}
\usepackage{multirow}
\newcommand{\myRule}{:$-$}
\newtheorem{dep}{Dependency}
\newtheorem{ppty}{Property}
\newtheorem{cor}{Corollary}
\newtheorem{pgm}{Program}
\newtheorem{ex}{Example}
\newenvironment{bt}{$\bf }{$}
\newcolumntype{L}{>{\centering\arraybackslash}m{3cm}}

\def\BibTeX{{\rm B\kern-.05em{\sc i\kern-.025em b}\kern-.08em
    T\kern-.1667em\lower.7ex\hbox{E}\kern-.125emX}}

\usepackage{booktabs} % For formal tables

\begin{document}

\title{IPAW 2020 Preprint: Efficient Computation of Provenance for Query Result Exploration}
\titlerunning{Efficient Computation of Provenance for Query Result Exploration}

\author{Murali Mani \and Naveenkumar Singaraj \and
Zhenyan Liu
}

\authorrunning{Murali Mani et al.}
\tocauthor{Murali Mani, Naveenkumar Singaraj, Zhenyan Liu}
\institute{University of Michigan Flint, Flint, Michigan, 48502
\email{mmani,nsingara,zhenyanl@umich.edu}}

\maketitle              % typeset the header of the contribution
\begin{abstract} 
Users typically interact with a database by asking queries and examining the results. We refer to the user examining the query results and asking follow-up questions as {\em query result exploration}. Our work builds on two decades of provenance research useful for {\em query result exploration}. Three approaches for computing provenance have been described in the literature: lazy, eager, and hybrid.
%identified different possible semantics for query result exploration, including {\em which}, {\em why}, {\em where}, and {\em how}. 
We investigate lazy and eager approaches that utilize constraints that we have identified in the context of query result exploration, as well as novel hybrid approaches. For the TPC-H benchmark, these constraints are applicable to 19 out of the 22 queries, and result in a better performance for all queries that have a join.
%but one query Q1 (Q1 has no joins and existing approaches perform as good as our approach). 
Furthermore, the performance benefits from our approaches are significant, sometimes several orders of magnitude.

\end{abstract}

\input{intro}

\input{background}

\input{singleRule}

\input{materialization}

\input{evaluation}

\section{Related Work} \label{sec:related}

Different provenance semantics as described in~\cite{FTD:CCT:09,PODS:GT:17} can be used for query result exploration. Lineage, or {\em which}-provenance~\cite{TODS:CWW:00} specifies which rows from the different input tables produced the selected rows in the result. 
%One of the properties of {\em which}-provenance is that it is "complete"~\cite{FTD:CCT:09}. Further,~\cite{FTD:CCT:09} mentions that {\em which}-provenance is invariant for equivalent queries with no self-joins. Actually, {\em which}-provenance is invariant even for equivalent queries with self-joins, provided different names are given for the same table, and these names are "consistent" across query rewritings (as would happen in typical optimization). 
{\em why}-provenance~\cite{ICDT:BKT:01} provides more detailed explanation than {\em which}-provenance and collects the input tuples separately for different derivations of an output tuple. 
%While {\em why}-provenance is not invariant for equivalent queries, a variant of {\em why}-provenance called {\em minimal witness basis} that consists of minimal elements of the {\em why}-provenance, is invariant for equivalent queries. 
{\em how}-provenance~\cite{PODS:GKT:07,FTD:CCT:09,PODS:GT:17} provides even more detailed information than {\em why}-provenance and specifies how the different input table rows combined to produce the result rows. Trio~\cite{VLDBJ:BSH+:08} provides a provenance semantics similar to {\em how}-provenance as studied in~\cite{FTD:CCT:09}.
%It has been noted that {\em how}-provenance is not invariant for equivalent queries~\cite{FTD:CCT:09}. 
Deriving different provenance semantics from other provenance semantics is studied in~\cite{FTD:CCT:09,PODS:GT:17}: {\em how}-provenance provides the most general semantics and can be used to compute other provenance semantics~\cite{FTD:CCT:09}. A hierarchy of provenance semirings that shows how to compute different provenance semantics is explained in~\cite{PODS:GT:17}. 
%Other provenance semantics that provide different kinds of explanations are studied as {\em where}-provenance~\cite{ICDT:BKT:01}. 
Another provenance semantics in literature is {\em where}-provenance~\cite{ICDT:BKT:01}, which only says where the result data is copied from. Provenance of non-answers studies why expected rows are not present in the result and is studied in~\cite{VLDBJ:LLG:19,SIGMOD:CJ:09,PVLDB:HCDN:08}. Explaining results using properties of the data are studied in~\cite{PVLDB:ROS:15,PVLDB:WM:13}.

For our work, we choose {\em which}-provenance even though it provides less details than {\em why} and {\em how} provenance because: (a) {\em which}-provenance is defined for queries with aggregate and group by operators~\cite{PODS:GT:17} that we study in this paper, (b) {\em which}-provenance is complete~\cite{FTD:CCT:09}, in that all the other provenance semantics provide explanations that only include the input table rows selected by {\em which}-provenance. As part of our future work, we are investigating computing other provenance semantics starting from {\em which}-provenance and the original user query, (c) {\em which}-provenance is invariant under equivalent queries (provided tables in self-joins have different and "consistent" names), thus supporting correlated queries (d) results of {\em which}-provenance is a set of tables that can be represented in the relational model without using additional features as needed by {\em how}-provenance, or a large number of rows as needed by {\em why}-provenance.

When we materialize data for query result exploration, the size of the materialized data can be an issue as identified by~\cite{PODS:GT:17}. Eager approaches record annotations (materialized data) which are propagated as part of provenance computation~\cite{VLDBJ:BCTV:05}. A hybrid approach that uses materialized data for computing provenance in data warehouse scenario as in~\cite{TODS:CWW:00} is studied in~\cite{DMDW:CW:00}. In our work, we materialize the results of some of the intermediate steps (views). While materializing the results of an intermediate step, we augment the result with the keys of some of the base tables used in that step. Note that the non-key columns are not stored, and the keys for all the tables may not need to be stored; instead, we selectively choose the base tables whose keys are stored based on the expected benefit and cost, and based on other factors such as workload.

%In our work, we identify that the complete row from a base table need not be materialized; instead only the key values of a row need to be materialized (annotating results with identifiers from base tables can be considered as materializing keys). Further, in our scenario, it is not required to materialize rows from every base table; instead, we can selectively choose which base tables to materialize based on the expected benefit and cost, and based on other factors such as workload. 

Other scenarios have been considered. For instance, provenance of non-answers are considered in~\cite{SIGMOD:CJ:09,PVLDB:HCDN:08}. In~\cite{VLDBJ:LLG:19}, the authors study a unified approach for provenance of answers and non-answers. However, as noted in~\cite{PODS:GT:17}, research on negation in provenance has so far resulted in divergent approaches. Another scenario considered is explaining results using properties of the data~\cite{PVLDB:ROS:15,PVLDB:WM:13}.

Optimizing queries in the presence of constraints has long been studied in database literature, including chase algorithm for minimizing joins~\cite{BOOK:AHV:95}. Join minimization in SQL systems has typically considered key-foreign key joins~\cite{DZone:Eder:17}. Optimization specific to provenance queries is studied in~\cite{CoRR:NKG+:18}. Here the authors study heuristic and cost based optimization for provenance computation. The constraints we study in this paper are tuple generating dependencies as will occur in scenario of query result exploration; these are more general than key-foreign key constraints. We develop practical polynomial time algorithms for join minimization in the presence of these constraints. 
%Inference of constraints is studied in~\cite{CoRR:NKG+:18,TODS:Klug:80}. For our work, we do limited inference of constraints, and infer the key in the presence of a group by clause in SQL select statements.

\section{Conclusions and Future Work} \label{sec:conc}

In this paper, we studied dependencies that are applicable to query result exploration. These dependencies can be used to optimize query performance during query result exploration. For the TPC-H benchmark, we could optimize the performance of 36.84\% (7 out of 19) of the queries that we considered. Furthermore, we investigated how additional data can be materialized and then be used for optimizing the performance during query result exploration. Such materialization of data can optimize the performance of query result exploration for almost all the queries.

One of the main avenues worth exploring is extensions to the query language that we considered.
%There are several avenues of work worth exploring. 
The dependencies we considered can be used when the body of a rule is a conjunction of predicates. We did not consider union queries, negation or outer joins. These will be interesting to explore as the dependencies do not extend in a straightforward manner. Another interesting future direction is studying effective ways of navigating the search space of possible materializations. Also, it will be worthwhile investigating how to start from provenance tables and define other provenance semantics (such as {\em how}-provenance) in terms of the provenance tables.

\bibliographystyle{splncs04}
\bibliography{provenance}

\end{document}

%% file: intro.tex
\section{Introduction} \label{sec:intro}

\begin{figure}[!ht]
\caption{User asks original query and gets results. Now the user explores these results.}
\centering
\includegraphics[height=1.3531in,width=3in]{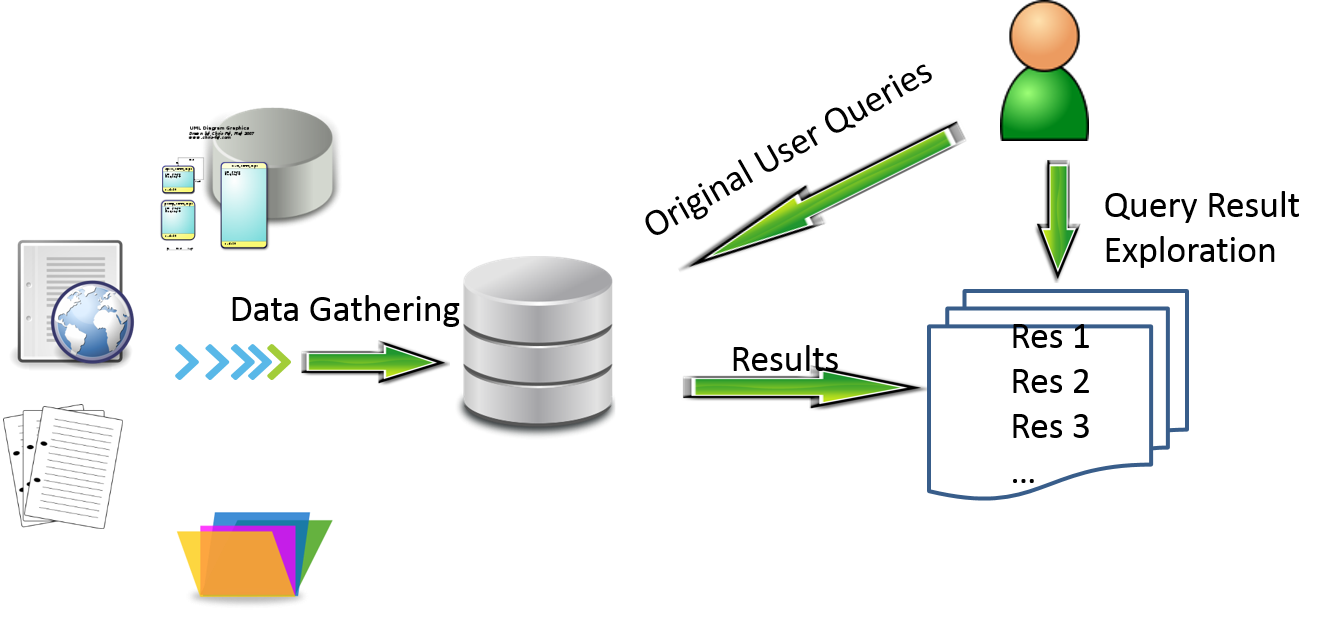}
\label{fig:scenario}
\end{figure}

Consider a user interacting with a database. Figure~\ref{fig:scenario} shows a typical interaction. Here the database is first assembled from various data sources (some databases might have a much simpler process, or a much more complex process). A user asks an {\em original query} and gets results. Now the user wants to {\em drill} deeper into the results and find out explanations for the results. We refer to this drilling deeper into the results as {\em query result exploration}.

For query result exploration, the user selects one or more interesting rows from the results obtained for the original user query, and asks questions such as: why are these rows in the result. The system responds by showing the rows in the tables that combined to produce those results the user is interested in.
Different provenance semantics as described in~\cite{FTD:CCT:09,PODS:GT:17} can be used for query result exploration. In this paper, we use the {\em which}-provenance semantics (also referred to as lineage) as in~\cite{TODS:CWW:00} and richer semantics is not needed. See Section~\ref{sec:related} for a discussion of different provenance semantics.

\begin{table}
\caption{Running Example: Tables (simplified) from TPC-H schema and sample data}\label{table:ex}
\begin{center}
\begin{tabular}{l@{\hskip 25pt}l@{\hskip 25pt}l}
\begin{tabular}{l}
\begin{bt}Customers\end{bt} \\
%$(\underline{c\_key}, c\_name, c\_address)$ \\ 

\begin{tabular}{|l|l|l|}
\hline
$\underline{c\_key}$ & $c\_name$ & $c\_address$ \\[2pt]
%\rowfont{\tiny} x & & \\
%\tiny{} & \tiny{} & \tiny{} \\
\hline
c1 & n1 & a1 \\
\hline
\end{tabular} 
\end{tabular}

&

\begin{tabular}{l}
\begin{bt}Orders\end{bt} \\
%$(\underline{o\_key}, c\_key, o\_date)$ \\ \\

\begin{tabular}{|l|l|l|}
\hline
$\underline{o\_key}$ & $c\_key$ & $o\_date$ \\[2pt]
\hline
o1 & c1 & d1 \\
o2 & c1 & d2 \\
\hline
\end{tabular} 
\end{tabular}

&

\begin{tabular}{l}
\begin{bt}Lineitem\end{bt} \\
%$(\underline{o\_key, linenum}, qty)$ \\ \\

\begin{tabular}{|l|l|l|}

\hline
$\underline{o\_key}$ & $\underline{linenum}$ & $qty$  \\[2pt]
\hline
o1 & l1 & 200 \\
o1 & l2 & 150 \\
o2 & l1 & 100 \\
o2 & l2 & 160 \\
\hline
\end{tabular}
\end{tabular}
\end{tabular}
\end{center}
\end{table}

\begin{ex}
\label{ex:q18}
\normalfont
Consider three tables from TPC-H~\cite{Web:TPCH} simplified and with sample data as shown in Table~\ref{table:ex}. Consider $Q18$ from TPC-H modified as in~\cite{Web:Jia:09} and simplified for our example. See that the query is defined in~\cite{Web:Jia:09} in two steps: first a view $Q18\_tmp$ is defined, which is then used to define the original query as view $R$. The results of these two views are also shown. \\

\begin{tabular}{ll}
\begin{tabular}{|p{2.7in}|}
\hline
(find total quantity for each order) \\
SQL: CREATE VIEW $Q18\_tmp$ AS \\
\hspace*{0.1in} SELECT $o\_key, sum(qty)$ as $t\_sum\_qty$ \\
\hspace*{0.1in} FROM \begin{bt}Lineitem\end{bt} \\
\hspace*{0.1in} GROUP BY $o\_key$ \\
\hline
\end{tabular}
&

\begin{tabular}{@{\hskip 30pt}c}
$Q18\_tmp$ \\
\begin{tabular}{|l|l|}
\hline
$o\_key$ & $t\_sum\_qty$ \\
\hline
o1 & 350 \\
o2 & 260 \\
\hline
\end{tabular}
\end{tabular}

\\

\begin{tabular}{|p{2.7in}|}
\hline
(for each order where total quantity is greater than 300, return
the customer and order information, as well as the
total quantity) \\
SQL: CREATE VIEW $R$ AS \\
\hspace*{0.1in}SELECT $c\_name, c\_key, o\_key, o\_date$, \\
\hspace*{0.5in}$sum(qty)$ as $tot\_qty$ \\
\hspace*{0.1in}FROM \begin{bt}Customers\end{bt}  NATURAL JOIN \begin{bt}Orders\end{bt} \\
\hspace*{0.5in}NATURAL JOIN \begin{bt}Lineitem\end{bt} \\
\hspace*{0.5in}NATURAL JOIN $Q18\_tmp$ \\
\hspace*{0.1in}WHERE $t\_sum\_qty > 300$\\
\hspace*{0.1in}GROUP BY  $c\_name$, $c\_key$, $o\_key$, $o\_date$ \\
\hline
\end{tabular}

&

\begin{tabular}{c}
$R$ \\
\begin{tabular}{|l|l|l|l|l|}
\hline
$c\_name$ & $c\_key$ &  $o\_key$ &  $o\_date$ & $tot\_qty$  \\
\hline
n1 & c1 & o1 & d1 & 350 \\
\hline
\end{tabular}
\end{tabular}
\end{tabular}
%\hfill{$\square$}
\end{ex}
\hfill{$\square$}

For this simplified example, there is one row in the result $R$. Suppose the user picks that row and wants to explore that row further. Suppose the user wants to find out what row(s) in the table \begin{bt}Customers\end{bt} produced that row. We use \begin{bt}R'\end{bt} to denote the table consisting of the rows picked by the user for query result exploration. We refer to the row(s) in the \begin{bt}Customers\end{bt} table that produced the row(s) in \begin{bt}R'\end{bt} as the provenance of \begin{bt}R'\end{bt} for the \begin{bt}Customers\end{bt} table, and denote it as $PCustomers$. In~\cite{TODS:CWW:00}, the authors come up with a query for determining this provenance shown below. Note that we sometimes use SQL syntax that is not valid, but intuitive and easier. \\

\noindent
\begin{tabular}{l@{\hskip 15pt}l}
\begin{tabular}{|p{3in}|}
\hline
SELECT \begin{bt}Customers\end{bt}.* \\
FROM \begin{bt}R'\end{bt} NATURAL JOIN \begin{bt}Customers\end{bt}  \\
\hspace*{0.1in}NATURAL JOIN \begin{bt}Orders\end{bt} NATURAL JOIN \\ 
\hspace*{0.1in}\begin{bt}Lineitem\end{bt} NATURAL JOIN $Q18\_tmp$ \\
WHERE $t\_sum\_qty > 300$\\
\hline
\end{tabular}
&
\noindent
\begin{tabular}{c}
$PCustomers$ \\
\begin{tabular}{|l|l|l|}
\hline
$c\_key$ & $c\_name$ & $c\_address$ \\
\hline
c1 & n1 & a1 \\
\hline
\end{tabular} 
\end{tabular}
\end{tabular} \\

However, if we observe closely, we can note the following. Given that the row in \begin{bt}R'\end{bt} appeared in the result of the original query with the value for $c\_key$ column as $c1$, and given that the key for \begin{bt}Customers\end{bt} is $c\_key$, the row from \begin{bt}Customers\end{bt} table that produced that row in $R$ must have $c\_key = c1$. Therefore the provenance retrieval query can be simplified as shown below. In this paper (Section~\ref{sec:singleRule}), we study such optimization of provenance retrieval queries formally. 

\noindent
\begin{center}
\begin{tabular}{|p{4.1in}|}
\hline
SELECT \begin{bt}Customers\end{bt}.*
FROM \begin{bt}R'\end{bt} NATURAL JOIN \begin{bt}Customers\end{bt} \\
\hline
\end{tabular}
\end{center}

As another example, consider the provenance of \begin{bt}R'\end{bt} in the inner \begin{bt}LineItem\end{bt} table (used for defining $Q18\_tmp$). This is computed in two steps. First we need to compute $PQ18\_tmp$. Below, we show the $PQ18\_tmp$ query as in~\cite{TODS:CWW:00}, and then our optimized $PQ18\_tmp$ query (using the same reasoning as for $PCustomers$).\\

\noindent
\begin{tabular}{l@{\hskip 45pt}l}
\begin{tabular}{|p{3in}|}
\hline
CREATE VIEW $PQ18\_tmp$ AS \\
SELECT $Q18\_tmp$.* \\
FROM \begin{bt}R'\end{bt} NATURAL JOIN \begin{bt}Customers\end{bt} \\
\hspace*{0.1in}NATURAL JOIN \begin{bt}Orders\end{bt} NATURAL JOIN \\ 
\hspace*{0.1in}\begin{bt}Lineitem\end{bt} NATURAL JOIN $Q18\_tmp$ \\
WHERE $t\_sum\_qty > 300$\\
\hline
\end{tabular}
&
\noindent
\begin{tabular}{c}
$PQ18\_tmp$ \\
\begin{tabular}{|l|l|}
\hline
$o\_key$ & $t\_sum\_qty$ \\
\hline
o1 & 350 \\
\hline
\end{tabular} 
\end{tabular}

\end{tabular} \\

\noindent
\begin{center}
\begin{tabular}{|p{3.8in}|}
\hline
CREATE VIEW $PQ18\_tmp$ AS \\
SELECT $Q18\_tmp$.* 
FROM \begin{bt}R'\end{bt} NATURAL JOIN $Q18\_tmp$ \\
\hline
\end{tabular} 
\end{center}

Now, the provenance of \begin{bt}R'\end{bt} in the inner \begin{bt}LineItem\end{bt} table can be computed using the following provenance retrieval query. \\

\noindent
\begin{tabular}{l@{\hskip 25pt}l}
\begin{tabular}{|p{3in}|}
\hline
SELECT LineItem.* \\
FROM \begin{bt}LineItem\end{bt} NATURAL JOIN $PQ18\_tmp$ \\
\hline
\end{tabular}
&
\begin{tabular}{l}
$PLineitem$ \\
\begin{tabular}{|l|l|l|}
\hline
$o\_key$ & $linenum$ & $qty$ \\
\hline
o1 & l1 & 200 \\
o1 & l2 & 150 \\
\hline
\end{tabular}
\end{tabular}
\end{tabular} \\

It is possible to further improve the performance of the above provenance retrieval query if we materialize some additional data. 
%For instance, consider the lazy provenance evaluation (with no materialization) studied in~\cite{TPC:GMA:13}. Suppose we modified this approach in~\cite{TPC:GMA:13} and materialized the "provenance table" during original user query execution. This materializes every column from every base table for every row in the result of the original user query. However, we can get good performance even when we materialize fewer additional data. 
Let us materialize the rows in $R$, along with the corresponding key value(s) from the inner \begin{bt}LineItem\end{bt} table for each row in $R$. We denote this result table augmented with additional keys and materialized as \begin{bt}RK\end{bt}. This will be done as follows.

% here

\noindent
\begin{tabular}{l@{\hskip 15pt}l}
\begin{tabular}{|p{2.8in}|}
\hline
CREATE VIEW $Q18\_tmpK$ AS \\
SELECT $Q18\_tmp$.*, \\
\hspace*{0.1in}\begin{bt}LineItem\end{bt}$.linenum$ AS linenum2 \\
FROM $Q18\_tmp$ NATURAL JOIN \begin{bt}LineItem\end{bt} \\
\hline
\end{tabular}
&
\begin{tabular}{c}
$Q18\_tmpK$ \\
\begin{tabular}{|l|l|l|}
\hline
$o\_key$ & $t\_sum\_qty$ & $linenum2$ \\
\hline
o1 & 350 & l1 \\
o1 & 350 & l2 \\
o2 & 260 & l1 \\
o2 & 260 & l2 \\
\hline
\end{tabular} 
\end{tabular}

\end{tabular} \\ \\

\noindent
\begin{tabular}{l@{\hskip 10pt}l}
\begin{tabular}{|p{1.8in}|}
\hline
CREATE TABLE \begin{bt}RK\end{bt} AS \\
SELECT $R$.*, linenum2 \\
FROM $R$ NATURAL JOIN \\
\hspace*{0.1in}$Q18\_tmpK$ \\
\hline
\end{tabular}
&
\begin{tabular}{c}
\begin{bt}RK\end{bt} \\
\begin{tabular}{|l|l|l|l|l|l|}
\hline
$c\_name$ & $c\_key$ & $o\_key$ & $o\_date$ & $tot\_qty$ & $linenum2$ \\
\hline
n1 & c1 & o1 & d1 & 350 & l1 \\
n1 & c1 & o1 & d1 & 350 & l2 \\
\hline
\end{tabular} 
\end{tabular}

\end{tabular} \\ \\

For this example, only the $linenum$ column needs to be added to the columns in $R$ as part of this materialization, because $o\_key$ is already present in $R$ (renamed as $linenum2$ to prevent incorrect natural joins). Now the provenance retrieval query for the inner \begin{bt}LineItem\end{bt} table can be defined as follows. \\

% we dont need PLineitem table here

\noindent
\begin{tabular}{l@{\hskip 10pt}l}
\begin{tabular}{|p{1.8in}|}
\hline
CREATE VIEW $RK'$ AS \\
SELECT * \\
FROM \begin{bt}R'\end{bt} NATURAL JOIN \\
\hspace*{0.1in}\begin{bt}RK\end{bt} \\
\hline
\end{tabular}
&
\begin{tabular}{c}
$RK'$ \\
\begin{tabular}{|l|l|l|l|l|l|}
\hline
$c\_name$ & $c\_key$ & $o\_key$ & $o\_date$ & $tot\_qty$ & $linenum2$ \\
\hline
n1 & c1 & o1 & d1 & 350 & l1 \\
n1 & c1 & o1 & d1 & 350 & l2 \\
\hline
\end{tabular} 
\end{tabular}

\end{tabular}

\noindent
\begin{center}
\begin{tabular}{|p{4.1in}|}
\hline
SELECT \begin{bt}LineItem\end{bt}.* 
FROM $RK'$ NATURAL JOIN \begin{bt}LineItem\end{bt} \\
\hline
\end{tabular}  
\end{center}

See that the provenance retrieval query for the \begin{bt}LineItem\end{bt} table in the inner block is now a join of 3 tables: \begin{bt}R'\end{bt}, \begin{bt}RK\end{bt} and \begin{bt}LineItem\end{bt}. Without materialization, the provenance retrieval query involved three joins also: \begin{bt}R'\end{bt}, $Q18\_tmp$ and \begin{bt}LineItem\end{bt}; however, $Q18\_tmp$ was a view. Our experimental studies confirm the huge performance benefit from this materialization.

Our contributions in this paper include the following:
\begin{itemize}
    \item We investigate constraints implied in our query result exploration scenario (Section~\ref{sec:dependencies}).
    \item We investigate optimization of provenance retrieval queries using the constraints. We present our results as a Theorem and we develop an Algorithm based on our theorem (Section~\ref{sec:singleRule}).
    \item We investigate materialization of select additional data, and investigate novel hybrid approaches for computing provenance that utilize the constraints and the materialized data 
    %that further help optimize provenance retrieval queries 
    (Section~\ref{sec:multiRule}).
    \item We perform a detailed performance evaluation comparing our approaches and existing approaches using TPC-H benchmark~\cite{Web:TPCH} and report the results (Section~\ref{sec:eval}).
\end{itemize}

% \noindent
% {\bf Outline:} The rest of the paper is organized as follows. Section~\ref{sec:background} introduces some of our notations, the set of queries supported as original user queries, algorithmic definition of provenance and the constraints for our scenario of query result exploration. Section~\ref{sec:singleRule} investigates the optimization of provenance retrieval queries using the constraints from Section~\ref{sec:background}, and without materialization. Materialization of additional data that further help optimize provenance retrieval queries is studied in Section~\ref{sec:multiRule}. Our experimental studies and results are described in Section~\ref{sec:eval}. Related work is discussed in Section~\ref{sec:related} and Section~\ref{sec:conc} concludes the work.

%% file: background.tex
\section{Preliminaries} \label{sec:background}

%\subsection{Notations for tables and views} \label{sec:notations}

%The notations for tables and views that we use in this paper are shown below. 
We use the following notations in this paper: a base table is in bold as \begin{bt}T_i\end{bt}, a materialized view is also in bold as \begin{bt}V_i\end{bt}, a virtual view is in italics as $V_i$. The set of attributes of table \begin{bt}T_i\end{bt}/materialized view \begin{bt}V_i\end{bt}/virtual view $V_i$ is $A_{T_i}$/$A_{V_i}$/$A_{V_i}$; the key for table \begin{bt}T_i\end{bt} is $K_i$. When the distinction between base table or virtual/materialized view is not important, we use $X_i$ to denote the table/view; attributes of $X_i$ are denoted $A_{X_i}$; the key (if defined) is denoted as $K_i$.

\subsection{Query Language} \label{sec:ql}

For our work, we consider SQL queries restricted to ASPJ queries and use set semantics. We do not consider set operators, including union and negation, or outer joins. We believe that extension to bag semantics should be fairly straightforward. However, the optimizations that we consider in this paper are not immediately applicable to unions and outer joins. Extensions to bag semantics, and these additional operators will be investigated in future work. For convenience, we use a Datalog syntax (intuitively extended with group by similar to relational algebra) for representing queries. We consider two types of rules (referred to as SPJ Rule and ASPJ Rule that correspond to SPJ and ASPJ view definitions in~\cite{TODS:CWW:00}) that can appear in the original query as shown in Table~\ref{table:rules}. 
%Note that Datalog~\cite{BOOK:ZCF:97} typically does not define group by, however our extended Datalog syntax (SPJA rule) mimics group by in relational algebra and should be intuitive. 
A query can consist of one or more rules. Every rule must be safe~\cite{BOOK:ZCF:97}. Note that Souffle\footnote{https://souffle-lang.github.io/} extends datalog with group by. In Souffle, our ASPJ rule will be written as two rules: an SPJ rule and a second rule with the group by. We chose our extension of Datalog (that mimics relational algebra) in this paper for convenience.
\begin{table*}
\caption{The two types of rules that can appear in original queries and their Datalog representation. For the ASPJ rule, $GL$ refers to the list of group by columns and $AL$ refers to the list of aggregations.}

\centering
\begin{tabular}{|l|l|}
\hline
SPJ Rule: &
$R (A_R)$ \myRule $X_1 (A_{X_1}), X_2 (A_{X_2}), \ldots, X_n (A_{X_n}).$ \\
\hline
ASPJ Rule: &
$R (GL, AL)$ \myRule $X_1 (A_{X_1}), X_2 (A_{X_2}), \ldots, X_n (A_{X_n}).$
\\
\hline
\end{tabular}
\label{table:rules}
\end{table*}

\begin{ex}\label{ex:q18detailed}
\normalfont
Consider query $Q18$ from TPC-H (simplified) shown in Example~\ref{ex:q18} written in Datalog. See that the two rules in $Q18$ are ASPJ rules, where the second ASPJ rule uses the $Q18\_temp$ view defined in the first ASPJ rule. The second rule can be rewritten as an SPJ rule; however, we kept it as an ASPJ rule as the ASPJ rule reflects the TPC-H query faithfully as is also provided in~\cite{Web:Jia:09}. \\

\noindent
\begin{tabular}{|m{\textwidth}|}
\hline
$Q18\_tmp(o\_key, sum(qty)$ as $t\_sum\_qty)$ \myRule \begin{bt}Lineitem\end{bt}. \\
$R$($c\_name$, $c\_key$, $o\_key$, $o\_date$, $sum(qty)$ as $tot\_qty$) \myRule \begin{bt}Customers\end{bt}, \begin{bt}Orders\end{bt}, \\ \hspace*{2.4in}\begin{bt}Lineitem\end{bt}, $Q18\_tmp$, $t\_sum\_qty > 300.$ \\
\hline
\end{tabular}
\end{ex}
\hfill{$\square$}

\subsection{Provenance Definition} \label{sec:provDef}

As said before, we use the {\em which}-provenance definition of~\cite{TODS:CWW:00}. 
%Note that this definition is similar to the {\em why}-provenance definition in~\cite{ICDT:BKT:01} and to the PI-CS definition in~\cite{TPC:GMA:13}, except for one difference: different provenance tables are defined in~\cite{TODS:CWW:00}, whereas the derivations of each result tuple is defined in~\cite{ICDT:BKT:01, TPC:GMA:13}. 
In this section, we provide a simple algorithmic definition for provenance based on our rules. 

The two types of rules in our program are both of the form: $R (A_R)$ \myRule $RHS$. We will use $A_{RHS}$ to indicate the union of all the attributes in the relations in $RHS$. For any rule, $R (A_R)$ \myRule $RHS$, the provenance for \begin{bt}R^\prime\end{bt} $\subseteq R$ in a table/view $X_i (A_{X_i}) \in RHS$ (that is, the rows in $X_i$ that contribute to the results \begin{bt}R^\prime\end{bt}) is given by the program shown in Table~\ref{table:provDefn}. See that $PView$ corresponds to the relational representation of {\em why}-provenance in~\cite{TPC:GMA:13}.

\begin{table*}
\caption{Algorithmic Definition of Provenance}
\begin{tabular}{|m{4.8in}|}
\hline
Algorithmic definition of provenance for rule: $R (A_R)$ \myRule $RHS$. The rows in table/view $X_i (A_{X_i}) \in RHS$ 
that contribute to \begin{bt}R^\prime\end{bt} $\subseteq R$ are represented as $PX_i$. \\ \\
$PView (A_R \cup A_{RHS})$ \myRule  $R (A_R), RHS.$ \\
$PX_i (A_{X_i})$ \myRule $PView,$ \begin{bt}R^\prime\end{bt}$(A_R).$ 
%\\ \\
%The above program for computing $PX_i(A_{X_i})$ can be written as a single rule as: \\
%$PX_i (A_{X_i})$ \myRule $R (A_R), RHS,$ \begin{bt}R^\prime\end{bt}$(A_R).$ 
\\
\hline
\end{tabular}
\label{table:provDefn}
\end{table*}

\begin{ex}
\normalfont
These examples are based on the schema and sample data in Table~\ref{table:ex}, and the $Q18\_tmp$ and $R$ views in Example~\ref{ex:q18detailed}. 

\noindent
\begin{tabular}{|m{4.8in}|}
\hline
Consider the definition of view $Q18\_tmp$ in Example~\ref{ex:q18detailed}; rows in the view $Q18\_tmp = \{(o1, 350), (o2, 260)\}$. Let rows selected to determine provenance $Q18\_tmp^\prime = \{(o2, 260)\}$. \\
\\
First $PView$ ($o\_key$,$t\_sum\_qty$, $linenum$, $qty$) is calculated as in Table~\ref{table:provDefn}. Here, $PView$ has four rows: \{ (o1, 350, l1, 200), (o1, 350, l2, 150), (o2, 260, l1, 100), (o2, 260, l2, 160)\} \\
Now $PLineItem$ is calculated (according to Table~\ref{table:provDefn}) as: \\
$PLineItem(o\_key, linenum, qty)$ \myRule $PView, Q18\_tmp^\prime.$ \\
The resulting rows for $PLineItem$ = \{ (o2, l1, 100), (o2, l2, 160)\} \\
\hline
\end{tabular}

% need to fix this
\hfill{$\square$}

\label{ex:provEx}
\end{ex}

%Examples of using the provenance definition in Table~\ref{table:provDefn} are shown in Table~\ref{table:provEx}. These examples are based on the schema and sample data in Table~\ref{table:ex}, and the $Q18\_temp$ and $R$ views in Table~\ref{table:q18detailed}. 

%As mentioned earlier, {\em which}-provenance is invariant for equivalent queries~\cite{FTD:CCT:09}. This is especially useful, because we do not need to consider correlated subqueries as they can be decorrelated. For instance our provenance definition gives the same results for the original TPC-H queries in~\cite{Web:TPCH} and for the decorrelated queries in~\cite{Web:Jia:09}.

\subsection{Dependencies} \label{sec:dependencies}

We will now examine some constraints for our query result exploration scenario that help optimize provenance retrieval queries. As in Section~\ref{sec:provDef}, the original query is of the form $R (A_R)$ \myRule $RHS$; and $A_{RHS}$ indicates the union of all the attributes in the relations in $RHS$. Furthermore, \begin{bt}R'\end{bt} $\subseteq R$.
We express the constraints as tuple generating dependencies below. While these dependencies are quite straightforward, they lead to significant optimization of provenance computation as we will see in later sections. 
%To check: Dependency \ref{dep:res} is full, tgd, single-head. Not sure if single-head matters or not.. not sure if we even need to check typed or not. Dependency  \ref{dep:join} and \ref{dep:inferred} are embedded, tgd, multi-head.

\begin{dep} \label{dep:res}
$\forall A_R,$ \begin{bt}R'\end{bt}$(A_R) \rightarrow R(A_R)$
\end{dep}

Dependency~\ref{dep:res} is obvious as the rows for which we compute the provenance, \begin{bt}R'\end{bt} is such that \begin{bt}R'\end{bt} $\subseteq R$.
For the remaining dependencies, consider $RHS$ as the join of the tables $X_1 (A_{X_1}), X_2 (A_{X_2}), \ldots, X_n (A_{X_n})$, as shown in Table~\ref{table:rules}.

%This is obvious as in our scenario, the rows for which we compute the provenance, \begin{bt}R'\end{bt} is such that \begin{bt}R'\end{bt} $\subseteq R$. Therefore Dependency~\ref{dep:res} is true for any \begin{bt}R'\end{bt} selected. 
%for our scenario as $R^prime$ is selected from the rows in $R$.

%For the remaining dependencies, consider $RHS$ as the join of the tables $X_1 (A_{X_1}), X_2 (A_{X_2}), \ldots, X_n (A_{X_n})$, as shown in Table~\ref{table:rules}.

\begin{dep} \label{dep:join}
$\forall A_R, \ R(A_R) \rightarrow \exists (A_{RHS} - A_R),$ $X_1 (A_{X_1}),$ $X_2 (A_{X_2}), \ldots, X_n (A_{X_n})$
\end{dep}

Dependency~\ref{dep:join} applies to both the rule types shown in Table~\ref{table:rules}. As any row in $R$ is produced by the join of $X_1 (A_{X_1}), X_2 (A_{X_2}), \ldots, X_n (A_{X_n})$, Dependency~\ref{dep:join} is also obvious. 
From Dependencies~\ref{dep:res} and \ref{dep:join}, we can infer the following dependency.

\begin{dep} \label{dep:inferred}
$\forall A_R,$ \begin{bt}R'\end{bt} $(A_R) \rightarrow \exists (A_{RHS} - A_R),$ $X_1 (A_{X_1}),$
$X_2 (A_{X_2}), \ldots, X_n (A_{X_n})$
\end{dep}

%% file: singleRule.tex
\section{Optimizing Provenance Queries without materialization} \label{sec:singleRule}

Consider the query for computing provenance given in Table~\ref{table:provDefn} after composition: $PX_i (A_{X_i})$ \myRule $R (A_R), RHS,$ \begin{bt}R^\prime\end{bt}$(A_R).$  Using Dependency~\ref{dep:res}, one of the joins in the query for computing provenance can immediately be removed. The program for computing provenance of \begin{bt}R'\end{bt} $\subseteq R$ in table/view $X_i$ is given by the following program. See that $X_i$ can be a base table or a view.
%This program is given by: (Equivalence can be shown trivially -- ask if you believe claim is not right, and I will try to see what is happening.) \\

\begin{pgm} \label{pgm:singleRule}
$PX_i(A_i)$ \myRule \begin{bt}R'\end{bt}$(A_R), RHS$.
\end{pgm}

Program~\ref{pgm:singleRule} is used by~\cite{TODS:CWW:00} for computing provenance. However, we will optimize Program~\ref{pgm:singleRule} further using the dependencies in Section~\ref{sec:dependencies}. Let $P_1$ below indicate the query in Program~\ref{pgm:singleRule}. Consider another query $P_2$ (which has potentially fewer joins than $P_1$). Theorem~\ref{thm:main} states when $P_1$ is equivalent to $P_2$. The proof uses the dependencies in Section~\ref{sec:dependencies} and is omitted. \\

\noindent $P_1: PX_i(A_{X_i})$ \myRule \begin{bt}R'\end{bt}$(A_R),$ $X_1 (A_{X_1}), X_2 (A_{X_2}), \ldots, X_n (A_{X_n})$. \\
\noindent $P_2: PX_i(A_{X_i})$ \myRule \begin{bt}R'\end{bt}, $X_{j_1} (A_{X_{j_1}}), X_{j_2} (A_{X_{j_2}}), \ldots, X_{j_q} (A_{X_{j_q}})$., \\
\hspace*{1.5in}where $\{j_1, j_2, \ldots, j_q\}$ $\subseteq$ $\{1, 2, \ldots, n\}$ \\

\noindent
{\bf Notation.} For convenience, we introduce two notations below. $A_{RHS}' = A_{X_{j_1}} \cup A_{X_{j_2}} \cup \ldots \cup A_{X_{j_q}}$. 
%In other words, $A_{RHS}^\prime$ denotes all the attributes in the RHS of $P_2$, not considering table \begin{bt}R'\end{bt}. 
Consider the tables that are present in the RHS of $P_1$, but not in the RHS of $P_2$. $A_{RHS}^{\prime\prime}$ denotes all the attributes in these tables. %In other words, $A_{RHS}^{\prime\prime} = \bigcup A_i$, $i \in \{1, 2, \ldots, n\} - \{j_1, j_2, \ldots, j_q\}$.

\begin{comment}
\begin{ppty} \label{prop:varName}
For ease of proof (w.l.o.g) we will name the variables the same as the name of the columns whenever possible. In fact, for both $P_1$ and $P_2$, we do not have to introduce any new variable names. We will therefore find that every predicate in the right side of $P_2$ will be present in the right side of $P_1$.
\end{ppty}
\end{comment}

\begin{theorem} \label{thm:main}
Queries $P_1$ and $P_2$ are equivalent, if 
%given the following condition. We will use $ASrc'$ to denote $ A_{j_1} \cup A_{j_2} \cup \ldots \cup A_{j_q}$, and $ASrc''$ to denote $\bigcup A_i$, $i \in \{1, 2, \ldots, n\} - \{j_1, j_2, \ldots, j_q\}$ \\
for every column $C \in A_{RHS}'$, at least one of the following is true:
\begin{itemize}
\item $A_R \rightarrow C$ (that is, $A_R$ functionally determines $C$)
\item $C \notin A_{RHS}''$ 
%(that is, $C$ is not present in any of the tables in $P_1$ that is not in $P_2$)
\end{itemize}
\end{theorem}

\begin{comment}
\noindent{\bf Illustration of Theorem~\ref{thm:main}}. \label{sec:singleRuleEx}
Consider the SPJA rule for $R$ for Q18 in TPC-H from Section~\ref{sec:ql}. The program for computing provenance $PCustomers$ before optimization and after optimization using Theorem~\ref{thm:main} are given below. See that \\
$A_{RHS}' = \{c\_key, c\_name, c\_address\}$; \\
$A_{RHS}'' = \{o\_key, c\_key, o\_date, linenum, qty, t\_sum\_qty\}$; \\
$A_R = \{c\_name, c\_key, o\_key, o\_date, total\_qty\}$. 
\\

\begin{minipage}[l]{6in}
\noindent $P_1$: $PCustomers (c\_key, c\_name, c\_address)$ \myRule \begin{bt}R'\end{bt},\begin{bt}Customers\end{bt},\\ \hspace*{0.3in}\begin{bt}Orders\end{bt}, \begin{bt}Lineitem\end{bt}, 
$Q18\_tmp, t\_sum\_qty > 300.$ \\
\noindent $P_2$: $PCustomers (c\_key, c\_name, c\_address)$ \myRule  \begin{bt}R'\end{bt}, \begin{bt}Customers\end{bt}. \\
\end{minipage}

There are three columns in $A_{RHS}'$. As $c\_key, c\_name \in A_R$, $A_R \rightarrow c\_key$, and $A_R \rightarrow c\_name$. For the column $c\_address$, we see that $c\_address \notin A_{RHS}''$. Therefore as per Theorem~\ref{thm:main}, $P_1$ and $P_2$ are equivalent queries. (See that the column $c\_name \notin A_{RHS}''$, and $A_R \rightarrow c\_address$ as well). See that Theorem~\ref{thm:main} resulted in program P2 with much fewer joins than the program P1; P2 is expected to perform better in practice.

%\noindent {\bf End of Thm Illustration}
\end{comment}

Based on Theorem \ref{thm:main}, we can infer the following corollaries. Corollary~\ref{cor:subset} says that if all the columns of $X_i$ are present in the result, no join is needed to compute the provenance of $X_i$. 
%This is true because if a row was in the result, then a corresponding row is present in $X_i$ with the same values for the shared columns and that forms the provenance. 
Corollary~\ref{cor:keys} says that if a key of $X_i$ is present in the result, then the provenance of $X_i$ can be computed by joining \begin{bt}R'\end{bt} and $X_i$. 
%This is true because if a key value for $X_i$ is present in the result, then the row in $X_i$ corresponding to that key value is the provenance for any picked result row.

\begin{cor} \label{cor:subset}
If $A_{X_i} \subseteq A_R$, then $PX_i(A_{X_i})$ \myRule \begin{bt}R'\end{bt}$(AR).$ 
\end{cor}

\begin{cor} \label{cor:keys}
If $K_i \subseteq A_R$, then $PX_i(A_{X_i})$ \myRule \begin{bt}R'\end{bt}$(AR)$, $X_i(A_{X_i}).$ 
\end{cor}

\subsection{Provenence Query Optimization Algorithm}

\begin{comment}
Let us come up with an algorithm to construct an efficient provenance query for any user query. (WHY DO WE WANT TO COME UP WITH AN ALGORITHM RATHER THAN USE AN ALGORITHM SIMPLISTICALLY ON CHASE USING THE DEPENDENCIES?? A NAIVE ALGORITHM WILL BE, CONSIDER ALL SUBSETS OF THE RHS, AND SEE WHICH ONES ARE EQUIVALENT USING CHASE.. (ANY OTHER BETTER WAY??) THE NUMBER OF QUERIES WE CONSIDER COULD BE EXPONENTIAL IN THE NUMBER OF JOINS IN THE QUERY. (WILL BE EXPONENTIAL IN SIZE OF QUERY UNLESS WE CAN PRUNE INTELLIGENTLY -- CAN WE PRUNE INTELLIGENTLY???. ALSO, CHASE USING THE DEPENDENCIES THAT WE CONSIDERED BEFORE, DO THEY GUARANTEE CONVERGENCE (OR) CAN IT BECOME UNDECIDABLE -- ARE THEY THE SAME?? OUR ALGORITHM IS POLYNOMIAL IN THE SIZE OF THE QUERY). 
\end{comment}

%Theorem~\ref{thm:main} describes when a provenance retrieval query with fewer joins is equivalent to the original provenance retrieval query (as in Program~\ref{pgm:singleRule}). 
In this section, we will come with an algorithm based on Theorem~\ref{thm:main} that starts with the original provenance retrieval query and comes up with a new optimized provenance retrieval query with fewer joins.
Suppose the original user query is: $R(A_R)$ \myRule $X_1 (A_{X_1})$, $X_2 (A_{X_2})$, $\ldots$, $X_n (A_{X_n})$. The user wants to determine the rows in $X_i$ that contributed to the results \begin{bt}R'\end{bt}$(A_R) \subseteq R(A_R)$. Note that $X_i$ can either be a base table or a view.

\begin{algorithm}
\caption{Efficient Provenance Retrieval Query} \label{alg:singleRule}
\begin{algorithmic}[1]
\State start with $CurRHS =$ \begin{bt}R'\end{bt}$(A_R)$
\If {$A_{X_i} \subseteq A_R$} \Return $CurRHS$
\EndIf
\State add $X_i$ to $CurRHS$
\State let $CurRHSTables = X_i$; $A_{RHS}'= \bigcup A_{X_j}$, where $X_j \in CurRHSTables$
\State let $RemTables$ = $\{X_1, X_2, \ldots, X_n\}$ -$X_i$; $A_{RHS}'' = \bigcup A_{X_j}$, where $X_j \in RemTables$
\While {there is a column $C \in A_{RHS}' \ \cap \ A_{RHS}''$, and there is no functional dependency $A_R \rightarrow C$} 
\State Add all tables in $RemTables$ that have the column $C$ to $CurRHS$, and to $CurRHSTables$. Adjust $A_{RHS}'$, $RemTables$, $A_{RHS}''$ appropriately.
\EndWhile
\State \Return $CurRHS$
\end{algorithmic}
\end{algorithm}

%\noindent
%{\bf Proof of Correctness of Algorithm~\ref{alg:singleRule}:} 
%In the above algorithm, if there is a column $C \in A_{RHS}'$ (i.e., $C$ is in one of the predicates in $CurRHS$) such that $A_R \not \rightarrow C$ (i.e., $A_R$ does not functionally determine $C$), all the predicates in RHS where $C$ appears will be added to $CurRHS$ in Steps 6 and 7 (in other words, $C$ will no longer be in $A_{RHS}''$).

\noindent
{\bf {Illustration of Algorithm \ref{alg:singleRule}}}

\noindent Consider the SPJA rule for $R$ for Q18 in TPC-H from Example~\ref{ex:q18detailed}.

\noindent
$R(c\_name, c\_key, o\_key, o\_date, sum(qty)$ as $total\_qty)$ \myRule  \\
\hspace*{0.2in}\begin{bt}Customers\end{bt}, \begin{bt}Orders\end{bt}, \begin{bt}Lineitem\end{bt}, $Q18\_tmp, t\_sum\_qty > 300.$ 

\noindent 
Algorithm~\ref{alg:singleRule} produces the provenance retrieval query for \begin{bt}Customers\end{bt} as follows. After line 3, $CurRHS =$ \begin{bt}R'\end{bt}, \begin{bt}Customers\end{bt}. At line 6, $A_{RHS}' \cap A_{RHS}''$ = $\{c\_key\}$. As $c\_key \in A_R$, and $A_R \rightarrow c\_key$, no more tables are added to $CurRHS$.
Thus, the final provenance retrieval query is:
%Provenance retrieval queries for \begin{bt}Orders\end{bt}, \begin{bt}Lineitem\end{bt}, and $Q18\_tmp$ are similarly optimized.
%\noindent
$PCustomers$ \myRule \begin{bt}R'\end{bt}, \begin{bt}Customers\end{bt}.
%\hspace*{0.1in}($A_{RHS}' \cap A_{RHS}''$ = $\{c\_key\}$. As $c\_key \in A_R$, $A_R \rightarrow c\_key$)
%See Section~\ref{sec:intro} for the intutition behind this optimization. 
%We find that Algorithm~\ref{alg:singleRule} optimizes the provenance retrieval queries for \begin{bt}Orders\end{bt} and \begin{bt}Lineitem\end{bt} tables, and view $Q18\_tmp$ as well.

%\noindent
%$POrders$ \myRule \begin{bt}R'\end{bt}, \begin{bt}Orders\end{bt}. (the column in $A_{RHS}' \cap A_{RHS}''$ is $o\_key$. As $o\_key \in A_R$, $A_R \rightarrow o\_key$)

%\noindent
%$PLineItem$ \myRule \begin{bt}R'\end{bt}, \begin{bt}LineItem\end{bt}. (the column in $A_{RHS}' \cap A_{RHS}''$ is $o\_key$, and $A_R \rightarrow o\_key$)

%\noindent
%$PQ18\_tmp$ \myRule \begin{bt}R'\end{bt}, $Q18\_tmp$. (note that Algorithm~\ref{alg:singleRule} is applicable to the view $Q18\_tmp$ as well. There are two columns in $A_{RHS}' \cap A_{RHS}''$: $o\_key$ and $t\_sum\_qty$. From $Q18\_tmp$ definition, we know that $o\_key \rightarrow t\_sum\_qty$. As $A_R \rightarrow o\_key$, $A_R \rightarrow t\_sum\_qty$ as well.)

%% file: materialization.tex
\section{Optimizing Provenance Queries with materialization} 
\label{sec:multiRule}
%\section{Extending Algorithm with materialization} \label{sec:multiRule}

%\section{Optimization using materialization}

In Section~\ref{sec:singleRule}, we studied optimizing the provenance retrieval queries for the lazy approach, where no additional data is materialized. Eager and hybrid approaches materialize additional data. An eager approach could be to materialize $PView$ (defined in Table~\ref{table:provDefn}). However, $PView$ could be a very large table with several columns and rows of data. In this section, we investigate novel hybrid approaches that materialize much less additional data, and perform comparable to (and often times, even better than) the eager approach that materializes $PView$. The constraints identified in Section~\ref{sec:dependencies} are still applicable, and are used to decrease the joins in the provenance retrieval queries.

\begin{figure}[h!]
\centering
\caption{Query with two steps } 
\includegraphics[width=2.5in]{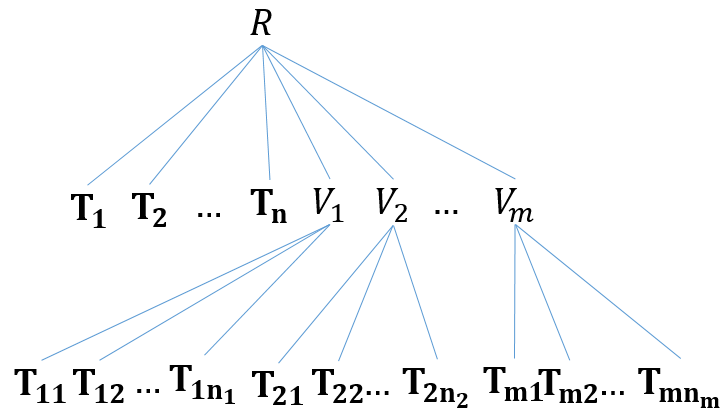} 
\label{fig:query}
\end{figure}

A user query can have multiple rules that form multiple steps (for instance, Q18 in TPC-H has two steps). While our results apply for queries with any number of steps, for simplicity of illustration, we consider only queries with two steps (the results extend in a straightforward manner to any number of steps). A query with two steps is shown in Figure~\ref{fig:query}. The Datalog program corresponding to Figure~\ref{fig:query} is shown in Program~\ref{pgm:oq}. 
$R$ is the result of the query. $R$ is defined using the base tables \begin{bt}T_1\end{bt}, \begin{bt}T_2\end{bt}, $\ldots$, \begin{bt}T_n\end{bt}, and the views $V_1$, $V_2$, $\ldots$, $V_m$. 
%In Figure~\ref{fig:query}, the views are shown as defined using only base tables. 
%(Each of the views can be defined using base tables or views.) 
Remember that from Section~\ref{sec:background}, \begin{bt}T_1\end{bt} has attributes $A_{T_1}$ and key attributes $K_1$;  \begin{bt}T_{1{n_1}}\end{bt} has attributes $A_{T_{1{n_1}}}$ and key attributes $K_{1{n_1}}$; $V_1$ has attributes $A_{V_1}$.

\noindent
\begin{minipage}{6in}
\begin{pgm}
\begin{tabular}{l}
$V_i (A_{V_i})$ \myRule \begin{bt}T_{i1}\end{bt}, \begin{bt}T_{i2}\end{bt}, \ldots, \begin{bt}T_{in_i}\end{bt}. 
\hspace*{0.3in}$\forall$ $i \in 1, 2, ..., m$ \\
$R (A_R)$ \myRule \begin{bt}T_1\end{bt}, \begin{bt}T_2\end{bt}, \ldots, \begin{bt}T_n\end{bt}, $V_1, V_2, \ldots, V_m$.
\end{tabular}
\label{pgm:oq}
\end{pgm}
\end{minipage} \\

%In such a case, while computing the results of the original user query, we consider materializing a view that includes all the rows in the original query result; for each row, this materialized view has all the columns in the original query and key columns for zero or more of the base tables used in the original query definition. 

Given a query $R$ as in Program~\ref{pgm:oq}, we materialize a view \begin{bt}RK\end{bt} with columns $A_{RK}$. $A_{RK}$ consists of the columns $A_R$ in $R$ and the keys of zero or more of the base tables used in $R$ (how $A_{RK}$ is determined is discussed later). 
%A base table \begin{bt}T\end{bt} is used in $R$ if $R$ is defined using $T$ or if $R$ is defined using a view $V$ that in turn uses \begin{bt}T\end{bt}. As an example, in 
%In Figure~\ref{fig:query}, two of the base tables used in $R$ are \begin{bt}T_1\end{bt} and \begin{bt}T_{11}\end{bt}. 
%For each view $V_i$ in Program~\ref{pgm:oq}, we will materialize keys for zero or more of the base tables. In addition, we will materialize keys for zero or more of the tables $T_j$, for $j \in 1, 2, \ldots, n$. In this case, we define \begin{bt}RK\end{bt} as follows. 
\begin{bt}RK\end{bt} is defined using $R$, the base tables that define $R$ and the $VK_i$ views corresponding to each of the $V_i$ that define $R$. See Program~\ref{pgm:oqmat}. $VK_i$ is a virtual view defined using $V_i$ and the tables that define $V_i$. If no keys are added to $V_i$ to form $VK_i$ (i.e., $A_{VK_i}$ = $A_{V_i}$), then $VK_i$ can be optimized to be just $V_i$. Algorithm~\ref{alg:singleRule} can be used to optimize $VK_i$ and \begin{bt}RK\end{bt} as well; details are omitted. \\

%We define \begin{bt}RK\end{bt} as follows. See that $VK_i$ is a virtual view. Definition of $VK_i$, \begin{bt}RK\end{bt} and how we rewrite the original user query ($OQ$) to use the materialized view \begin{bt}RK\end{bt} are shown in Program~\ref{pgm:oqmat}. 
%If we do not materialize the keys for any of the base tables used in $V_i$, then $VK_i$ will end up being the same as $V_i$. 

\noindent
\begin{minipage}{3.25in}
\begin{pgm}
\begin{tabular}{l}
$V_i(A_{V_i})$ \myRule \begin{bt}T_{i1}\end{bt}, \begin{bt}T_{i2}\end{bt}, \ldots, \begin{bt}T_{in_i}\end{bt}.
\hspace*{0.3in}$\forall$ $i \in 1, 2, ..., m$ \\
$R(A_R)$ \myRule \begin{bt}T_1\end{bt}, \begin{bt}T_2\end{bt}, \ldots, \begin{bt}T_n\end{bt}, $V_1, V_2, \ldots, V_m$. \\ \\
$VK_i(A_{VK_i})$ \myRule $V_i$, \begin{bt}T_{i1}\end{bt}, \begin{bt}T_{i2}\end{bt}, \ldots, \begin{bt}T_{in_i}\end{bt}.
\hspace*{0.01in}$\forall$ $i \in 1, 2, ..., m$ \\
\begin{bt}RK\end{bt}$(A_{RK})$ \myRule $R$, \begin{bt}T_1\end{bt}, \begin{bt}T_2\end{bt}, \ldots, \begin{bt}T_n\end{bt}, $VK_1$, $VK_2$, $\ldots$, $VK_m$. \\
$OQ (A_R)$ \myRule \begin{bt}RK\end{bt}. \\
\end{tabular} 
\label{pgm:oqmat}
\end{pgm}
\end{minipage} \\

The original user query results (computed as $R$ in Program~\ref{pgm:oq}) are computed by $OQ$ in Program~\ref{pgm:oqmat}. This is because we assume that \begin{bt}RK\end{bt} is materialized during the original user query execution and we expect that computing $OQ$ from \begin{bt}RK\end{bt} will be faster than computing the results of $R$.

For query result exploration, suppose that the user selects \begin{bt}R'\end{bt} $\subseteq R$ and wants to find the provenance of \begin{bt}R'\end{bt} in the table \begin{bt}T_i\end{bt}. We will assume that \begin{bt}T_i\end{bt} is a base table that defines $V_j$. For this, we first define $RK' \subseteq$ \begin{bt}RK\end{bt} as shown below. \\

\noindent
$RK'$ \myRule \begin{bt}R'\end{bt},\begin{bt}RK\end{bt}. \\

$RK'$ denotes the rows in \begin{bt}RK\end{bt} corresponding to the rows in \begin{bt}R'\end{bt}. Now to compute the provenance of \begin{bt}R'\end{bt} in the table \begin{bt}T_i\end{bt}, we compute the provenance of $RK'$ in the table \begin{bt}T_i\end{bt}. There are two cases:

\begin{pgm} \label{pgm:provmulti}
\begin{tabular}{l@{\hskip 10pt}l@{\hskip 25pt}l}
Case 1: & $K_i \subseteq A_{RK}$: & $PT_i$ \myRule $RK'$, \begin{bt}T_i\end{bt}. \\

Case 2: & $K_i \nsubseteq A_{RK}$:  & $PT_i$ \myRule $PV_j, V_{jRHS}$. \\ 
\multicolumn{3}{c}{($V_{jRHS}$ is the RHS of the rule that defines $V_j$.)}
\end{tabular}
\end{pgm}

Case 1 is similar to Corollary~\ref{cor:keys} except that $R$ may not be defined using \begin{bt}T_i\end{bt} directly. 
%The proof of correctness follows from Theorem~\ref{thm:main}.
For Case 2, $V_j$ is defined using \begin{bt}T_i\end{bt} directly. $PV_j$ is the provenance of $RK'$ in the view $V_j$, computed recursively using Program~\ref{pgm:provmulti}. Given $PV_j$, the rule for computing the provenance of $PV_j$ in the table \begin{bt}T_i\end{bt} is given by Program~\ref{pgm:singleRule}. 
Both the rules in Program~\ref{pgm:provmulti} can be optimized using Algorithm~\ref{alg:singleRule}.

\begin{ex}
\normalfont
\label{ex:materialize}

Consider $Q18$ from Example~\ref{ex:q18detailed}. There are 4 base tables used in $Q18$ -- \begin{bt}Customers\end{bt}, \begin{bt}Orders\end{bt}, \begin{bt}Lineitem1\end{bt} and \begin{bt}Lineitem2\end{bt}. We distinguish the two \begin{bt}Lineitem\end{bt} tables; \begin{bt}Lineitem2\end{bt} denotes the table used in $Q18\_tmp$ definition.

The materialized \begin{bt}RK\end{bt} view contains the columns in $R$ and additionally the key for \begin{bt}Lineitem2\end{bt} table. The key for the \begin{bt}Lineitem2\end{bt} table is $(o\_key, linenum)$; however $o\_key$ is already present in $R$. Therefore only the $linenum$ column from \begin{bt}Lineitem2\end{bt} is added in $A_{RK}$. The revised program (as in Program~\ref{pgm:oqmat}) that materializes \begin{bt}RK\end{bt} and computes $OQ$ is shown below. Note that optimizations as in Algorithm~\ref{alg:singleRule} are applicable (for example, definition of \begin{bt}RK\end{bt}); details are omitted. \\

%Let us assume that we materialize the keys for \begin{bt}Customers\end{bt} and \begin{bt}Lineitem2\end{bt} tables. 
%The revised program will look as follows. Note that $c\_key$ (key for the \begin{bt}Customers\end{bt} table) is already present in $R$. The key for the \begin{bt}Lineitem2\end{bt} table is $(o\_key, linenum)$; however $o\_key$ is already present in $R$. Therefore only the $linenum$ column from \begin{bt}Lineitem2\end{bt} is added in $A_{RK}$.) \\

\noindent
\begin{minipage}{\textwidth}
$Q18\_tmp(o\_key, sum(qty)$ as $t\_sum\_qty)$ \myRule \begin{bt}Lineitem\end{bt}. \\
$R(c\_name, c\_key, o\_key, o\_date, o\_totalprice, sum(qty)$ as $total\_qty)$ \\
\hspace*{0.2in}\myRule \begin{bt}Customers\end{bt}, \begin{bt}Orders\end{bt}, \begin{bt}Lineitem\end{bt}, $Q18\_tmp, t\_sum\_qty > 300.$ 
\end{minipage} \\ \\

\noindent
\begin{minipage}{\textwidth}
$Q18\_tmpK(o\_key, linenum$ as $linenum2, t\_sum\_qty)$ \myRule $Q18\_tmp$, \begin{bt}Lineitem\end{bt}. \\
\begin{bt}RK\end{bt}$(c\_name$, $c\_key$, $o\_key$, $o\_date$, $o\_totalprice$, $linenum2$,
$total\_qty)$ \\
\hspace*{0.2in}\myRule $R, Q18\_tmpK.$ \\ 
%(the rule for \begin{bt}RK\end{bt} is obtained after performing optimizations as in Algorithm~\ref{alg:singleRule}).
%\end{minipage} \\ \\
%\noindent
%\begin{minipage}{\textwidth}
$OQ(c\_name, c\_key, o\_key, o\_date, o\_totalprice, total\_qty)$ \myRule \begin{bt}RK\end{bt}. \\
\end{minipage} \\

\noindent
Let \begin{bt}R'\end{bt} denote the selected rows in $R$ whose provenance we want to explore. 
%First note that the keys for $Customers$, $Orders$ and $Lineitem2$ tables are materialized in $RK$. 
%The provenance for these tables will be computed simply as a join with the respective tables. As the key for $Lineitem1$ table is not materialized in $RK$, the provenance for this table will be more complex as will see below. \\
To compute their provenance, we first need to determine which rows in \begin{bt}RK\end{bt} correspond to the rows in \begin{bt}R'\end{bt}. This is done as: \\

\noindent
$RK' (A_{RK})$ \myRule \begin{bt}R'\end{bt}, \begin{bt}RK\end{bt}. \\

\noindent
Now, we need to compute the provenance of the rows in $RK'$ from the different tables, which is computed as follows. See that all the rules have been optimized using Algorithm~\ref{alg:singleRule}, and involve a join of $RK'$ and one base table.
\\

\noindent
\begin{minipage}{\textwidth}
$PCustomers (c\_key, c\_name, c\_address)$ \myRule $RK'$, \begin{bt}Customers\end{bt}. \\
$POrders(o\_key, c\_key, o\_date, o\_totalprice)$ \myRule $RK'$, \begin{bt}Orders\end{bt}. \\
$PLineitem1(o\_key, linenum, qty)$ \myRule $RK'$, \begin{bt}Lineitem\end{bt}. \\
$PLineitem2(o\_key, linenum, qty)$ \myRule 
$RK'$ $(c\_name$, $c\_key$, $o\_key$, \\
\hspace*{0.2in}$o\_date$, $o\_totalprice$, $linenum2$ as $linenum$, $total\_qty)$, \begin{bt}Lineitem\end{bt}.
\end{minipage}

%See that all the rules have been optimized using Algorithm~\ref{alg:singleRule}, and involve a join of $RK'$ and one base table.
%Further, we have inferred that $o\_key$ is the key for $Q18\_tmp$.
\hfill{$\square$}

\end{ex}

%----------------------------------------------------------------

\subsection{Determining the keys to be added to the materialized view}

When we materialize \begin{bt}RK\end{bt}, computing the results of the original user query is expected to take longer because we consider that materialization of \begin{bt}RK\end{bt} is done during original query execution, and because \begin{bt}RK\end{bt} is expected to be larger than the size of $R$: the number of rows (and the number of columns) in \begin{bt}RK\end{bt} will not be fewer than the number of rows (and the number of columns) in $R$. 
%We consider materializing \begin{bt}RK\end{bt} is done during original query execution. 
However, materialization typically benefits result exploration because the number of joins to compute the provenance for some of the base tables is expected to be smaller (although it is possible that the size of \begin{bt}RK\end{bt} might be large and this may slow down the provenance computation).

For the materialized view \begin{bt}RK\end{bt}, we consider adding keys of the different base tables and compute the cost .vs. benefit. The ratio of the estimated number of rows of \begin{bt}RK\end{bt} and the estimated number of rows in $R$ forms the cost. The ratio of the number of joins across all provenance computations of base tables with and without materialization give the benefit. We use a simple cost model that combines the cost and the benefit to determine the set of keys to be added to \begin{bt}RK\end{bt}. For the example query $Q18$, the provenance retrieval queries for \begin{bt}Customers\end{bt}, \begin{bt}Orders\end{bt} and \begin{bt}Lineitem\end{bt} tables in the outer block already involve only one join. Therefore no keys need to be added to improve the performance of these three provenance retrieval queries. However, we can improve the performance of the provenance retrieval query for the \begin{bt}Lineitem\end{bt} table in the inner block by adding the keys for the inner \begin{bt}Lineitem\end{bt} table to \begin{bt}RK\end{bt} as shown in Example \ref{ex:materialize}.

%We consider keys for different base tables to be added to the materialized view \begin{bt}RK\end{bt} and compute the cost .vs. benefit. The ratio of the estimated number of rows of \begin{bt}RK\end{bt} and the estimated number of rows in $R$ forms the cost. The ratio of the number of joins across all provenance computations of base tables with and without materialization give the benefit. We use a simple cost model that combines the cost and the benefit to find the set of keys to be materialized. For the example query $Q18$, the provenance retrieval queries for \begin{bt}Customers\end{bt}, \begin{bt}Orders\end{bt} and \begin{bt}Lineitem\end{bt} tables in the outer block already involve only one join. Therefore no keys need to be added to improve the performance of these three provenance retrieval queries. However, we can improve the performance of the provenance retrieval query for the \begin{bt}Lineitem\end{bt} table in the inner block by materializing the keys for the inner \begin{bt}Lineitem\end{bt} table as shown in Example \ref{ex:materialize}.

For \begin{bt}RK\end{bt}, we currently consider adding the key for every base table as part of the cost-benefit analysis. In other words, the number of different hybrid options we consider is exponential in the number of tables in the original user query. For each option, the cost .vs. benefit is estimated and one of the options is selected. As part of future work, we are investigating effective ways of searching this space. Other factors may be included in our cost model to determine which keys to be added to \begin{bt}RK\end{bt}, including the workload of provenance queries.

%% file: evaluation.tex
\section{Evaluation} \label{sec:eval}

For our evaluation, we used the TPC-H~\cite{Web:TPCH} benchmark. We generated data at 1GB scale. Our experiments were conducted on a PostgreSQL 10 database server running on Windows 7 Enterprise operating system. The hardware included a 4-core Intel Xeon 2.5 GHz Processor with 128 GB of RAM. For our queries, we again used the TPC-H benchmark.
%We studied 13 out of the 22 queries, as we omitted queries with outer joins and with exists. The queries we studied from the benchmark are: Q1, Q2, Q3, Q6, Q8, Q9, Q10, Q12, Q14, Q15, Q16, Q18, and Q19. 
The queries provided in the benchmark were considered the original user queries. Actually, we considered the version of the TPC-H queries provided by~\cite{Web:Jia:09}, which specifies values for the parameters for the TPC-H benchmark and also rewrites nested queries. For the result exploration part, we considered that the user would pick one row in the result of the original query (our solutions apply even if multiple rows were picked) and ask for the rows in the base tables that produce that resulting row.

We compare the following approaches: 
\begin{itemize}
\setlength\itemsep{-1pt}
    %\item The approach in~\cite{TODS:CWW:00} that we refer to as: W. We consider that in this approach no additional data is materialized (lazy approach). In other words, we do not consider the materialization studied in the hybrid approach in~\cite{DMDW:CW:00}.
    \item The approach in~\cite{TODS:CWW:00} that we refer to as: W (lazy approach). No additional data is materialized; the materialization studied in~\cite{DMDW:CW:00} is not considered.
    \item The approach in~\cite{TPC:GMA:13} that we refer to as: G. Here we assume that the relational representation of provenance is materialized while computing the original user query (eager approach). Provenance computation is then translated into mere look-ups in this materialized data.
    \item Algorithm~\ref{alg:singleRule} without materialization that we refer to as: O1 (lazy approach).
    \item Our approach with materialization from Section~\ref{sec:multiRule} that we refer to as: O2 (hybrid approach).
\end{itemize}

\subsection{Usefulness of our optimization rules}

Algorithm~\ref{alg:singleRule} results in queries with much fewer joins. We tested the provenance retrieval queries for $Q18$ from TPC-H as given in~\cite{Web:Jia:09} (for our experiments, the schema and the queries were not simplified as in our running example). The times observed are listed in Table~\ref{table:exp1}. See that the provenance retrieval queries generated by Algorithm~\ref{alg:singleRule} (O1) run much faster than the ones used in~\cite{TODS:CWW:00} (W).

\begin{table}[h!]
\caption{O1 compared to W for $Q18$ in~\cite{Web:Jia:09}. All times are reported in milliseconds.} \label{table:exp1}
\centering
\begin{tabular}{c|l|l|l|}
& $PCustomers$ & $POrders$ & $PLineItem$\\
\hline
O1 & 0.07 & 0.06 & 0.30 \\
\hline
W & 1522.44 & 1533.88 & 1532.74 \\
\hline
\end{tabular}
\end{table}

We considered all the TPC-H queries as given in~\cite{Web:Jia:09} except for the ones with outer joins (as we do not consider outer joins in this paper). Of the 22 TPC-H queries, the queries with outer joins are Q13, Q21, Q22, and these were not considered. $Q19$ has $or$ in its predicate, which can be rewritten as a union. However, we considered the $or$ predicate as a single predicate without breaking it into a union of multiple rules. For 7 out of these 19 queries, O1 results in provenance retrieval queries with fewer joins than the ones in W. They were Q2, Q3, Q7, Q10, Q11, Q15 and Q18. In other words, Algorithm~\ref{alg:singleRule} was useful for around 36.84\% of the TPC-H queries.

\subsection{Usefulness of materialization}

%We now studied the cost and benefits of materialization. For this, 
For $Q18$~\cite{Web:Jia:09}, we compared the time to compute the original query results (OQ) and the time to compute the provenance of the four tables for the four approaches: O1, W, G and O2. 
%For O2, 
%which is our approach without materialization; W is the approach in~\cite{TODS:CWW:00} without materialization; G is the approach in~\cite{ICDE:GA:09}; O2 is our approach with materialization where 
%our hybrid approach 
The materialized view \begin{bt}RK\end{bt} in O2 included the key for the \begin{bt}LineItem\end{bt} table in the inner block. The results are shown in Table~\ref{table:exp2}.

\begin{table}[h!]
\caption{Performance Benefits of materialization proposed in Section~\ref{sec:multiRule} for Q18 in~\cite{Web:Jia:09}. All times are reported in milliseconds.} 
\centering
\begin{tabular}{c|l|l|l|l|}
& O1 & W & G & O2 \\
\hline
OQ & 5095.67 & 5095.67 & 5735446.19 & 13794.26 \\
\hline
PCustomers & 0.07 & 1522.44 & 3.86 & 0.96 \\
\hline
POrders & 0.06 & 1533.88 & 3.73 & 0.43 \\
\hline
PLineItem1 & 0.30 & 1532.74 & 5.77 & 0.59 \\
\hline
PLineItem2 & 1641.52 & 1535.22 & 6.16 & 0.43 \\
\hline
\end{tabular}
\label{table:exp2}
\end{table}

There are several points worth observing in Table~\ref{table:exp2}. 
% When we compare our hybrid approach with materialization (O2) and the eager approach corresponding to~\cite{TPC:GMA:13} (G), we see that 
%We typically expects  O2 outperforms G in all cases. 
We typically expect O2 to outperform G in computing the results of the original user query. This is because G maintains all the columns of every base table in the materialized view, whereas O2 maintains only some key columns in the materialized view - in this case, the materialized view consists of the columns in $R$ and only one addition column $linenum2$. The performance impact of this is significant as G takes about 420 times the time taken by O2 to compute the results of the original user query. Actually the time taken by G is about 5700 seconds, which is likely to be unacceptable. On the other hand, O2 takes about 2.7 times the time taken by O1 for computing the results of the original user query. 
%We drilled deeper to find out for O2, whether the materialization of \begin{bt}RK\end{bt} was taking much time or computing the results of the original user query from the materialized view \begin{bt}RK\end{bt}. We actually 
Drilling down further, we found that computing the results from the materialized view \begin{bt}RK\end{bt} took about 0.39 milliseconds for O2 and about 3.07 milliseconds for G (Table~\ref{table:exp3}{\bf (b)}). 

\begin{table}[h!]
\caption{{\bf{(a)}} Comparing the size of the tables: R (result of the original user query), {\bf RK}\_G (materialized view {\bf RK} used by G) and {\bf RK}\_O2 (materialized view {\bf RK} used by O2). {\bf{(b)}} Comparing time for computing materialized view {\bf RK} and time for computing original query results from {\bf RK} for Q18~\cite{Web:Jia:09}. All times are reported in milliseconds.} 
\centering
\resizebox{\textwidth}{!}{%
\begin{tabular}{c@{\hskip 20pt}c}
\begin{tabular}{l|l|l|l|}
& R & {\bf RK}\_G & {\bf RK}\_O2 \\
\hline
\# Columns & 6 & 51 & 7 \\
\hline
\# Rows & 57 & 2793 & 399 \\
\hline
\end{tabular}
&
\begin{tabular}{r|l|l|}
& G & O2 \\
\hline
Computing {\bf RK} &  5735443.12 &  13793.88 \\
\hline
Computing OQ from {\bf RK} &  3.07 & 0.39 \\
\hline
\end{tabular} \\
{\bf{(a)}} & {\bf{(b)}}
\end{tabular}
}
\label{table:exp3}
\end{table}

%\begin{table}[h!]
%\caption{Comparing time for computing materialized view and time for computing original query results from the materialized view for O2 and G for Q18 in~\cite{Web:Jia:09}. All times are reported in milliseconds.}
%\centering
%\begin{tabular}{l|l|l|}
%& G & O2 \\
%\hline
%Computing MV &  5735443.12 &  13793.88 \\
%\hline
%Computing OQ from MV &  3.07 & 0.39 \\
%\hline
%\end{tabular}
% \label{table:exp4}
%\end{table}

We expect G to outperform O2 in computing the provenance. This is because the provenance retrieval in G requires a join of \begin{bt}R'\end{bt} with \begin{bt}RK\end{bt}. O2 requires a join of 3 tables (if the key is included in \begin{bt}RK\end{bt}). For Q18, the provenance retrieval query for $LineItem2$ requires a join of \begin{bt}R'\end{bt} with \begin{bt}RK\end{bt} to produce $RK'$, which is then joined with \begin{bt}LineItem\end{bt} table. However the larger size of \begin{bt}RK\end{bt} in G (Table~\ref{table:exp3}{\bf (a)}) results in O2 outperforming G for provenance retrieval (Table~\ref{table:exp2}). 

%We expect the provenance retrieval query for O2 
In practice, O2 will never perform worse than O1 for provenance retrieval.
%the provenance retrieval  will never perform worse than the provenance retrieval query for O1. 
This is because for any table, the provenance retrieval query for O1 (that does not use $RK'$, but instead uses \begin{bt}R'\end{bt}) may be used instead of the provenance retrieval query for O2 (that uses $RK'$ as in Program~\ref{pgm:provmulti}) if we expect the performance of the provenance retrieval query for O1 to be better. However, we have not considered this optimization in this paper.

Other things to note are that computing the results of the original query for $O1$ and $W$ is done exactly the same way. Moreover, for $Q18$, O1 outperforms all approaches even in provenance retrieval except for $PLineItem2$. This is because Algorithm~\ref{alg:singleRule} is able to optimize the provenance retrieval queries significantly for $PCustomers$, $POrders$, $PLineItem1$. However, for $PLineItem2$, the provenance retrieval required computing $PQ18\_tmp$ and then using it to compute $PLineItem2$, which needed more joins. Usually, we expect every provenance retrieval query from O1 to outperform W, but in this case W did outperform O1 for $PLineItem2$ (by a small amount); we believe the reason for this is the extra joins in W ended up being helpful for performance (which is not typical).

%After studying in detail the performance for one query, we compared how the different approaches perform for several TPC-H queries. 
We report on the 19 TPC-H queries without outer joins in Table~\ref{table:exp5}. In this table, OQ refers to the time taken for computing the results of the original user query, AP (average provenance) refers to the time taken to compute the provenance averaged over all the base tables used in the query, and MP (minimum provenance) refers to the minimum time to compute provenance over all the base tables used in the query. For W, we typically expect AP and MP to be almost the same (unless for nested queries); this is because in W, every provenance retrieval query (for non-nested original user queries) performs the same joins. Similarly for G, we typically expect AP and MP to be almost the same (because every provenance computation is just a look-up in the materialized data), except for the difference in the size of the results. For O1 and O2, MP might be significantly smaller than AP because some provenance computation might have been optimized extensively (example: Q2, Q10, Q11, Q15, Q18).

%\begin{longtable}[h!]
%\centering
%\begin{tabular}{cc|l|l|l|l|}
\begin{table}[h!]
\caption{Summary of experiments. The times are reported in milliseconds to two decimal places accuracy. However, considering the width of the table, if the time is 100 ms or greater, we report in scientific notation with two significant numbers.}
\centering
\resizebox{\textwidth}{!}{%
\begin{tabular}{c|lll|lll|lll|lll|}
&\multicolumn{3}{c|}{O1} &\multicolumn{3}{c|}{W} &\multicolumn{3}{c|}{G} &\multicolumn{3}{c|}{O2} \\
& OQ & AP & MP & OQ & AP & MP & OQ & AP & MP & OQ & AP & MP \\[0.5pt]
\hline
Q1 & 3.4e3 & 3.2e3 & 3.2e3 & 3.4e3 & 3.2e3 & 3.2e3 & 1.5e5 & 3.7e4 & 3.7e4 & 1.1e5 & 2.7e4 & 2.7e4 \\[0.5pt]
\hline
Q2 & 55.88 & 37.41 & 0.21 & 55.88 & 55.59 & 43.03 & 1.3e4 & 1.25 & 0.98 & 7.7e3 & 0.61 & 0.52\\[0.5pt]
\hline
Q3 & 8.7e2 & 0.06 & 0.04 & 8.7e2 & 0.09 & 0.08 & 2.9e3 & 45.57 & 43.11 & 2.5e3 & 4.28 & 3.47\\[0.5pt]
\hline
Q4 & 4.1e3 & 5.3e3 & 4.3e3 & 4.1e3 & 5.3e3 & 4.3e3 & 3.3e4 & 1.6e2 & 1.4e2 & 7.7e3 & 4.5e2 & 3.5e2\\[0.5pt]
\hline
Q5 & 6.3e2 & 6.7e2 & 6.5e2 & 6.3e2 & 6.7e2 & 6.5e2 & 2.9e3 & 13.01 & 10.71 & 2.5e3 & 11.45 & 4.73\\[0.5pt]
\hline
Q6 & 6.2e2 & 6.7e2 & 6.7e2 & 6.3e2 & 6.7e2 & 6.7e2 & 3.2e3 & 90.28 & 90.28 & 2.9e3 & 9.0e2 & 9.0e2\\[0.5pt]
\hline
Q7 & 8.7e2 & 6.7e2 & 6.6e2 & 8.7e2 & 6.7e2 & 6.6e2 & 4.9e3 & 13.22 & 11.12 & 4.5e3 & 12.26 & 6.88\\[0.5pt]
\hline
Q8 & 8.3e2 & 1.7e3 & 1.6e3 & 8.3e2 & 1.7e3 & 1.6e3 & 4.1e3 & 5.05 & 2.17 & 3.3e3 & 7.92 & 3.74
\\[0.5pt]
\hline
Q9 & 3.7e3 & 2.3e6 & 2.2e6 & 3.7e3 & 2.3e6 & 2.2e6 & 2.2e5  & 1.7e2 & 1.7e2 & 1.9e5 & 7.2e2 & 1.0e2
\\[0.5pt]
\hline
Q10 & 1.5e3 & 99.69 & 0.06 & 1.5e3 & 1.3e2 & 1.3e2 & 9.6e6 & 1.1e2 & 1.1e2 & 3.1e3 & 1.0e2 & 30.27 \\[0.5pt]
\hline
Q11 & 4.3e2 & 2.6e2 & 4.06 & 4.3e2 & 6.0e2 & 3.9e2 & 1.9e6 & 8.2e4 & 7.4e4  & 1.3e3 & 3.1e2 & 0.58 \\[0.5pt]
\hline
Q12 & 8.9e2 & 7.9e2 & 7.8e2 & 8.9e2 & 7.9e2 & 7.8e2 & 4.0e3 & 9.15 & 8.60 & 3.9e3 & 30.00  & 21.31\\[0.5pt]
\hline
Q14 & 7.7e2 & 9.8e2 & 9.2e2  & 7.7e2 & 9.8e2 & 9.2e2 & 4.3e3 & 1.8e2 & 1.7e2 & 3.3e3 & 5.8e2 & 2.8e2  \\[0.5pt]
\hline
Q15 & 1.43e3 & 1.0e3 & 4.62 & 1.4e3 & 2.2e3 & 1.4e3 & 2.0e5 & 6.0e4 & 3.0e4 & 1.7e5 & 9.7e4 & 5.5e4\\[0.5pt]
\hline
Q16 & 1.2e3 & 1.3e2 & 1.1e2 & 1.2e3 & 1.3e2 & 1.1e2 & 4.9e3 & 55.37 & 54.03 & 2.6e3 & 2.2e2 & 2.2e2 \\[0.5pt]
\hline
Q17 & 4.2e3 & 5.9e3 & 4.3e3 & 4.2e3 & 5.9e3 & 4.3e3 & 2.2e4 & 41.79 & 37.96 & 2.2e4 & 4.3e3 & 4.3e3\\[0.5pt]
\hline
Q18 & 5.1e3 & 4.1e2 & 0.06 & 5.1e3 & 1.5e3 & 1.5e3 & 5.7e6 & 4.88 & 3.73 & 1.4e4 & 0.60 & 0.43 \\[0.5pt]
\hline
Q19 & 2.4e3 & 2.4e3 & 2.4e3 & 2.4e3 & 2.4e3 & 2.4e3 & 1.3e4 & 13.35 & 12.62 & 1.3e4 & 86.23 & 83.44 \\[0.5pt]
\hline 
Q20 & 2.0e3 & 2.3e3 & 1.9e3 & 2.0e3 & 2.3e3 & 1.9e3 & 6.9e4 & 0.34 & 0.28 & 4.0e3 & 5.6e2 & 0.21 \\[0.5pt]
\hline 
\end{tabular}}
 \label{table:exp5}
\end{table}

We find that except for one single table query Q1, where W performs same as O1, our approaches improve performance for provenance computation, and hence for result exploration. Furthermore, the eager materialization approach (G) could result in prohibitively high times for original result computation.